\documentclass{elsarticle}

\usepackage{setspace,graphics,booktabs,color,enumerate,subfigure,url,bm}
\usepackage{amssymb}
\usepackage{mathrsfs}
\usepackage{multirow}
\usepackage{amsmath, amssymb, color}
\usepackage{graphicx}
\usepackage{subfigure}
\usepackage{algorithm}
\usepackage{algorithmic}
\definecolor{DarkBlue}{rgb}{0,0.08,0.45}
\usepackage[backref = false, bookmarks, colorlinks = true, plainpages = false, citecolor = DarkBlue , urlcolor = DarkBlue, filecolor = DarkBlue, linkcolor = DarkBlue]{hyperref}
\usepackage{array}
\usepackage{float}
\floatstyle{ruled}
\newfloat{algorithm}{htp}{lop}
\floatname{algorithm}{Algorithm}
\newtheorem{definition}{Definition}

\newtheorem{theorem}{Theorem}
\newtheorem{lemma}{Lemma}

\begin{document}




\begin{frontmatter}
\title{{Scheduling under Linear Constraints}}

\author[ts]{Kameng Nip}
\ead{njm13@mails.tsinghua.edu.cn}

\author[ts]{Zhenbo Wang\corref{cor}}
\ead{zwang@math.tsinghua.edu.cn}
\cortext[cor]{Corresponding author}

\author[mu]{Zizhuo Wang}
\ead{zwang@umn.edu}
\address[ts]{Department of Mathematical Sciences, Tsinghua University, Beijing, China}
\address[mu]{Department of Industrial and Systems Engineering, University of Minnesota, MN, USA}

\begin{abstract}
We introduce a parallel machine scheduling problem in
which the processing times of jobs are not given in advance but are
determined by a system of linear constraints. The objective is to
minimize the makespan, i.e., the maximum job completion time among
all feasible choices. This novel problem is motivated by various
real-world application scenarios. We discuss the computational
complexity and algorithms for various settings of this problem. In
particular, we show that if there is only one machine with an
arbitrary number of linear constraints, or there is an arbitrary
number of machines with no more than two linear constraints, or both
the number of machines and the number of linear constraints are
fixed constants, then the problem is polynomial-time solvable via
solving a series of linear programming problems. If both the number
of machines and the number of constraints are inputs of the problem
instance, then the problem is NP-Hard. We further propose several
approximation algorithms for the latter case.
\end{abstract}

\end{frontmatter}

\section{Introduction}
\label{sec:intro}

A scheduling problem aims to allocate resources to jobs, so as to
meet a specific objective, e.g., to minimize the makespan or the
total completion time. One common assumption in the classical
scheduling problem is that the processing times of jobs are
deterministic and are given in advance. However, in practice, the
processing times are usually uncertain/unknown or could be part of
the decisions. A number of works in the literature have proposed
various scheduling models in which the processing times are
uncertain/unknown, such as the stochastic scheduling problem
\citep{MRW84,MSU99,Dean05} and the robust scheduling problem
\citep{DK95,Kasperski05,KZ08}. In the stochastic scheduling problem,
it is assumed that the processing times are random variables and the
expected makespan is considered. In the robust scheduling problem,
it is assumed that the processing time of each job belongs to a
certain set and the objective is to find a robust schedule under
some performance criterion (e.g., minimize the maximum absolute
deviation of total completion time, or the total lateness). Note
that in either the stochastic or the robust scheduling problems, the
processing times are still exogenously given.

In the presented paper, we introduce a new scheduling model. In our
model, the processing times of jobs are not exogenously given,
instead they can be chosen as part of the decisions, but they must
satisfy a set of linear constraints. We call this problem the
``scheduling under linear constraints" (SLC) problem. Note that the
SLC problem reduces to the classical parallel machine scheduling
problem $P||C_{\max}$ when the processing times of jobs are given
(or equivalently, when the linear constraints have a unique
solution). This problem is related to the scheduling problem with
controllable processing times studied in the literature
\citep{NZ88,NZ90,SS07}. In the latter problem, the processing times
of jobs are controlled by factors such as the starting times and the
sequence of the jobs, while in our problem, the processing times are
part of the decision variables. The SLC problem is also related to
the lot sizing and scheduling problem in production planning, which
decides the type and amount of jobs to process at each time period
over a time horizon \citep{Drexl199573,Drexl1997221,Haase1994}.
However, in the SLC problem, each task must be completed in a
consecutive time interval, while in the lot sizing and scheduling
problem, an activity can be scheduled in multiple non-consecutive
periods. Furthermore, the objective of the lot sizing and scheduling
problem is to minimize the total costs, including the setup costs,
the inventory holding costs, etc, which is significantly different
from the objective in our problem \citep{POMS:POMS77}.

%

In the following, we provide a few examples that motivate the study
of the SLC problem.

\begin{enumerate}
\item \emph{Industrial Production Problem.} Perhaps the earliest
motivation for the scheduling problem came from manufacturing
\citep[e.g., see][]{Pinedo09,Pinedo12}. Suppose a manufacturer
requires certain amounts of different raw metals, and he needs to
extract them from several alloys. There are several machines that
can extract the alloys in parallel. We focus on the procedure of
extracting the alloys, of which the goal is to finish as early as
possible. In this problem, the processing times of extracting each
alloy depend on the processing quantities, and traditionally, they
are predetermined in advance. However, in practice, those quantities
are determined by the demands of the raw metals and can be solved as
a feasible solution to a blending problem \citep{Dan60,ES07}.
Sometimes, each alloy also has its own maximum quantity. An example
of such a scenario is given in Table \ref{tab:example1}.

\begin{table}
\begin{center}
\begin{tabular}{p{6em}<{\centering}|p{2em}<{\centering}p{2em}<{\centering}p{2em}<{\centering}p{2em}<{\centering}p{2em}<{\centering}|c}
                 \toprule
                  Composition& \multicolumn{5}{c|}{Alloy} & Demand\\
                     & 1 & 2 & 3 & $\cdots$ & $n$ & \\
                  \hline
                  iron &  24 & 8 & 3 & $\cdots$ & 2 & $\geq$ 56\\
                  copper &  3 & 3 & 3 & $\cdots$ & 1 & $\geq$ 30\\
                  $\vdots$ &$\vdots$&$\vdots$&$\vdots$&$\vdots$ & $\vdots$& $\vdots$  \\
                  aluminium & 4 & 33 & 137 & $\cdots$ & 100  & $\geq 1000$\\
                  \hline
                  Max. of alloy&  & &  &  &   & Quantity\\
                  \hline
                  1 &  1 & 0 & 0 & $\cdots$ & 0 & $\leq$ 10\\
                  2 &  0 & 1 & 0 & $\cdots$ & 0 & $\leq$ 7\\
                  3 & 0 & 0 & 1 & $\cdots$ & 0 & $\leq$ 20\\
                  $\vdots$ &$\vdots$&$\vdots$&$\vdots$&$\vdots$ & $\vdots$& $\vdots$  \\
                  $n$ & 0 & 0 & 0 & $\cdots$ & 1  & $\leq 15$\\
                 \bottomrule
\end{tabular}
\end{center}\caption{Example for the Industrial Production Problem}
\label{tab:example1}
\end{table}

In the example shown in Table \ref{tab:example1}, the demand of iron
is $56$, and each unit of alloy 1 contains 24 units of iron, each
unit of alloy 2 contains 8 units of iron, etc. Let $x_i$ be the
quantity of alloy $i$ to be extracted. Then the requirement on the
demand of iron can be represented as a linear inequality $24 x_1 + 8
x_2 + 3 x_3 + \cdots + 2 x_n \geq 56$. Furthermore, the maximum
amount of alloy 1 available is 10, which can be represented as a
linear inequality $x_1 \leq 10$. Similarly, we can write linear
constraints for the demand of other metals and the quantity for
other alloys. In this problem, the decision maker needs to determine
the nonnegative job quantities $x_1,\ldots,x_n$ satisfying the above
linear constraints, and then assign these jobs to the parallel
machines such that the last completion time is minimized. This
problem can be viewed as a minimum makespan parallel machine
scheduling problem, where the processing times of jobs satisfy some
linear constraints.\vspace{0.3cm}
%

\item \emph{Advertising Media Selection Problem.} A company has several
parallel broadcast platforms which can broadcast advertisements
simultaneously, such as multiple screens in a shopping mall or
different spots on a website. There is a customer who wants to
broadcast his advertisements (ad $1,\ldots,n$) on these
platforms.\footnote{This example can be easily extended to cases
with multiple customers.} It is required that each individual
advertisement must be broadcast without interruption and the running
time of each advertisement has to satisfy some linear constraints.
The company needs to decide the running times $x_i$ allocated to
each advertisement $i$, and also which advertisement should be
released on which platform as well as the releasing order. The
objective is to minimize the completion time. An example of such a
problem is given in Table \ref{tab:example2}.

\begin{table}
\begin{center}
\begin{tabular}{p{11em}<{\centering}|p{2.5em}<{\centering}p{2.5em}<{\centering}p{2.5em}<{\centering}p{2.5em}<{\centering}p{2.5em}<{\centering}|c}
                 \toprule
                  \multirow{2}{*}{sum of} & \multicolumn{5}{c|}{each unit time broadcast provides} & \\
                   & ad 1 & ad 2 & ad 3 & $\cdots$ & ad $n$ & \\
                  \hline
                  \small{attractions to women} & 20 & 100 & 100 & $\cdots$ & 10 & $\geq$ 500\\
                  \small{attractions to men} & 15 & 10 & 0 & $\cdots$ & 80 & $\geq$ 500\\
                  \small{attractions to teens} & 30 & 0 & 30 & $\cdots$ & 100 & $\geq$ 200\\
                  $\vdots$ &$\vdots$&$\vdots$&$\vdots$&$\vdots$ & $\vdots$& $\vdots$  \\
                  \small{max time for ad 1} & 1 & 0 & 0 & $\cdots$ & 0 & $\leq$ 20\\
                  \small{min time for ad 1} & 1 & 0 & 0 & $\cdots$ & 0 & $\geq$ 10\\
                  \small{max time for ad 2} & 0 & 1 & 0 & $\cdots$ & 0 & $\leq$ 35\\
                  $\vdots$ &$\vdots$&$\vdots$&$\vdots$&$\vdots$ & $\vdots$& $\vdots$  \\
                  \bottomrule
\end{tabular}
\end{center}\caption{Example for the Advertising Media Selection
Problem}\label{tab:example2}
\end{table}

Similar to the first example, the above-described problem can be
naturally formulated as a minimum makespan parallel machine
scheduling problem in which the parameters (running times of the
advertisements) are determined by a system of linear constraints.
\vspace{0.3cm}

\item \emph{Transportation Problem.} Both linear programming and machine scheduling problems have
extensive applications in the field of transportation management
\citep{ES07,Pinedo09,Pinedo12}. The parallel machine scheduling
problem has many similarities with the transportation scheduling
models. For example, a fleet of tankers or a number of workers can
be considered as a parallel machine environment, and transporting or
handling cargo is analogous to processing a job \citep{Pinedo09}.
Meanwhile, the transportation problem can be formulated as a linear
program. Let $x_{ij}$ be the capacity of cargo that needs to be
transported from origin $i$ to destination $j$. They often have to
satisfy certain supply and demand constraints, which are usually
linear constraints.

In practice, the decision maker decides how to assign cargo (jobs)
to tankers or workers (parallel processors), so as to finish the
handling as quickly as possible. This is a parallel machine
scheduling problem. And the processing times
usually depend on $x_{ij}$s, which have to satisfy some linear constraints as mentioned above. This also leads to a parallel machine scheduling problem with linear constraints. \\
\end{enumerate}

In this paper, we study the SLC problem, discussing the
computational complexity and algorithms for this problem under
various settings. In particular, we show that if there is only one
machine with an arbitrary number of linear constraints, or there is
an arbitrary number of machines with no more than two linear
constraints, or both the number of machines and the number of linear
constraints are fixed constants, then the problem is polynomial-time
solvable via solving a series of linear programming problems. If
both the number of machines and the number of constraints are inputs
of the problem instance, then the problem is NP-Hard. We further
propose several approximation algorithms for the latter case. We
summarize our results in Table \ref{tabs}. In Table \ref{tabs}, the
parameters $n$, $m$, $k$ stand for the number of jobs, machines and
constraints, respectively. The
upper line in each cell indicates the computational complexity of
the problem, where P refers to polynomial-time solvable and ? refers
to complexity unknown; the lower line indicates the running time if
the problem is polynomial-time solvable, or the performance ratios
of our approximation algorithms if it is NP-Hard. The superscripts
indicate the section where the corresponding result appears. The parameter $L$ is the input size of
the problem and $K$ is a value depending on $k$ and $m$ whose explicit expression will be given in Section \ref{sec:amak}.

\begin{table}[htbp]
\begin{center}
{\small\begin{tabular}{p{2.8em}<{\centering}p{3.8em}<{\centering}p{8em}<{\centering}p{8.2em}<{\centering}p{8em}<{\centering}}
                 \toprule
                  & $~~~k = 1~~~$ & $~~~k = 2~~~$ & $k
                   \geq3$ (fixed) & $k\geq3$ (input)\\
                  \toprule
                   \multirow{2}{*}{\small$m = 1$} &P & P & P & P \\
                  \cmidrule(l){2-5}
                   & $O(n)^{[\ref{sec:1k}]}$ & $O(n^{2}L)^{[\ref{sec:fmfk}]}$ & $O(n^3L)^{[\ref{sec:1m}]}$  & $O((n+k)^3L)^{[\ref{sec:1m}]}$ \\
                \toprule
                   {$m \geq 2$}  & P & P & P & NP-Hard\\
                    \cmidrule(l){2-5}
                    (fixed) & $O(n)^{[\ref{sec:1k}]}$ & {\footnotesize{$O(n^{\min\{m + 1, 4\}}L)^{[\ref{sec:fmfk},\ref{sec:am2k}]}$}} & $O(n^{m + k - 1}L)^{[\ref{sec:fmfk}]}$ & PTAS$^{[\ref{sec:fmak}]}$ \\
                   \toprule
                   {$m \geq 2$} &  P &  P & ? & Strongly NP-Hard\\
                    \cmidrule(l){2-5}
                    (input) & $O(n)^{[\ref{sec:1k}]}$ & $O(n^4L)^{[\ref{sec:am2k}]}$  & \multicolumn{2}{c}{$\min\{\frac{m}{m-K}, 2-\frac{1}{m}\}^{[\ref{sec:amak}]}$}\\
                  \bottomrule
\end{tabular}}
\caption{Summary of Results}\label{tabs}
\end{center}
\end{table}

One interesting conclusion from our result is that although parallel
machine scheduling is in general an intractable problem, a seemingly
more complicated problem --- parallel machine scheduling with linear
constraints --- can be simpler and tractable in many cases. This
suggests that instead of finding a feasible solution to the linear
constraints and then assigning it to the machines, a decision maker
should consider them jointly. In other words, it is often beneficial
to consider the problem with a big-picture perspective.

The remainder of the paper is organized as follows: In Section
\ref{sec_pro}, we formally state the problem studied in this paper
and briefly review some existing results. We study the simplest case
in which there is only one machine or one constraint in Section
\ref{sec:1m1k}. In Section \ref{sec:fm}, we consider the case with
at least two but still a fixed number of machines. In Section
\ref{sec:am}, we investigate the case where the number of machines
is an input of the instance. Finally, some concluding remarks are
provided in Section \ref{sec_end}.

\section{Problem Description}\label{sec_pro}
The scheduling problem under linear constraints is formally defined
as below:
\begin{definition}
Given $m$ identical machines and $n$ jobs. The processing times of
the jobs are nonnegative and satisfy $k$ linear inequalities. The
goal of the scheduling problem under linear constraints (SLC) is to
determine the processing times of the jobs such that they satisfy
the linear constraints and to assign the jobs to the machines to
minimize the makespan.
\end{definition}

Formally, let $x_i$ be the processing time of job $i$. The
processing times $\bm{x} = (x_1, \ldots , x_n)$ should satisfy
\begin{eqnarray}\label{eq:LP}
\begin{array}{ll}
A\bm{x} \geq \bm{b}, \quad \bm{x} \geq 0,
\end{array}
\end{eqnarray}
where $A \in \mathbb{R}^{k \times n}$ and $\bm{b}\in \mathbb{R}^{k
\times 1}$.

Parallel machine scheduling with the objective of minimizing the
makespan is one of the most basic models in various scheduling
problems \citep[e.g., see][]{ChPoWo98}. This problem is NP-Hard even
if there are only two machines, and it is strongly NP-Hard when the
number of machines is an input of the instance \citep{GJ79}. On the
algorithmic side, \cite{Graham66} proposed a $(2 -
\frac{1}{m})$-approximation algorithm for parallel machine
scheduling with $m$ machines. This method, known as the list
scheduling (LS) rule, is in fact one of the earliest approximation
algorithms. Later, \cite{Graham69} presented the longest processing
time (LPT) rule with an approximation ratio of $(\frac{4}{3} -
\frac{1}{3m})$ and a polynomial-time approximation scheme (PTAS)
when the number of machines is fixed. For the case of a fixed number
of machines, \cite{Sahni76} further proposed a fully polynomial-time
approximation scheme (FPTAS). When the number of machines is an
input, \cite{HS87} showed that a PTAS exists.

At first sight, the SLC problem can be formulated as the following
optimization problem:
\begin{eqnarray*}\hspace{0.5cm}
  \begin{array}{lcl}
    \min & t  \\
   s.t. & \sum_{j = 1}^{m} y_{ij} = 1 & \qquad\forall i = 1, \ldots , n\\
        & \sum_{i = 1}^{n} x_{i}y_{ij} \leq t & \qquad\forall j = 1, \ldots , m\\
        & A\bm{x} \geq \bm{b}\\
       & \bm{x}, t \geq 0\\
       & y_{ij} \in \{0, 1\} & \qquad\forall i, j,\\
   \end{array}
\end{eqnarray*}
where $y_{ij} = 1$ indicates that job $i$ is assigned to machine
$j$. This can be viewed as a nonconvex mixed integer (binary)
quadratic programming problem \citep{BL12,Koppe11} or a mixed
integer (binary) bilinear programming problem \citep{AS93,GAMCD13}.
In general, such problems are NP-Hard and extremely hard to solve.
In fact, it is unknown whether the mixed integer quadratic programming
problem lies in NP \citep{BL12,Jeroslow73}. However,
with the special structure of the problem, we will show that several
cases of the SLC problem can be solved in polynomial time or
approximated within a constant factor.

\section{Single Machine or Single Constraint}
\label{sec:1m1k}

\subsection{Single Machine}
\label{sec:1m}

If there is only one machine, then the classical parallel machine
scheduling problem becomes trivial since the makespan is simply the
total processing time. For the SLC problem, it is equivalent to
solving the following linear program:
\begin{eqnarray*}{\textrm{(LP1)}}\hspace{0.5cm}
\begin{array}{ll}
\min & \sum\limits_{i = 1}^n x_i  \\
s.t. & A\bm{x} \geq \bm{b} \\
     & \bm{x} \geq 0.
\end{array}
\end{eqnarray*}

Therefore, we have the following conclusion. We refer the readers to
\citet{Ye} for the complexity of the interior point methods.
\begin{theorem}
The SLC problem with a single machine can be solved in polynomial
time, in particular, in $O((n+k)^{3}L)$ time by the interior point
methods, where $L$ is the size of input length.
\end{theorem}

\subsection{Single Constraint}
\label{sec:1k}

In this subsection, we study the SLC problem with only one
constraint, that is, $k = 1$ and $A$ is a $1\times n$ matrix. In
this case, the linear constraints can be written as
\begin{eqnarray*}
\sum\limits_{i = 1}^n a_i x_i \geq b, \quad  \bm{x} \geq 0.
\end{eqnarray*}

Without loss of generality, we assume that $a_1 \geq a_2 \geq \cdots
\geq a_{n}$ and $b \geq 0$. If all $a_i$ are nonpositive, then this
problem is trivial (all $x_i = 0$ if $b = 0$, or infeasible if $b >
0$). Therefore, we assume that there is at least one $a_i > 0$. We
define $n'= \min\{\max\{i| a_i > 0\}, m\}$, where $m$ is the number
of machines, and $\sigma = \sum^{n'}_{i = 1}a_i$. We have the
following result:

\begin{theorem}\label{th:1k}
For the SLC problem with one constraint, the optimal decisions are
$x_1 = \cdots = x_{n'} = b/\sigma$ and $x_i  = 0$ otherwise, and the
optimal makespan is $b/\sigma$.
\end{theorem}

{\noindent\bf Proof.} Consider the following linear program:
\begin{eqnarray*}{\textrm{(LP2)}}\hspace{0.5cm}
\begin{array}{lcl}
\min & t  \\
s.t. & \sum\limits_{i = 1}^{n} a_i x_i \geq b\\
& \sum\limits_{i = 1}^{n} x_{i} \leq mt\\
& x_{i} \leq t & \qquad\forall i = 1, \ldots , n\\
& \bm{x}, t \geq 0.
\end{array}
\end{eqnarray*}

Note that (LP2) can be viewed as a relaxation of the SLC problem,
since any optimal solution to the SLC problem is feasible to (LP2)
by choosing $\bm{x}$ as the processing times and $t$ as its
makespan. Suppose we have an optimal solution $(\bm{x}, t)$ to
(LP2). If it is also feasible to the SLC problem, that is, the jobs
have processing times $\bm{x}$ and can be assigned to the $m$
machines with makespan at most $t$, then it must also be optimal to
the SLC problem.

The dual problem of (LP2) is
\begin{eqnarray*}{\textrm{(DP2)}}\hspace{0.5cm}
\begin{array}{lcl}
\max & bu  \\
s.t. & a_i u - y_i - v \leq 0 &\qquad \forall i = 1, \ldots , n\\
& \sum\limits_{i = 1}^{n} y_{i} + mv  \leq 1\\
& u, v, \bm{y} \geq 0.\\
\end{array}
\end{eqnarray*}

Let $x_i = b/\sigma$ for $i = 1, \ldots , n'$ and $x_i  = 0$ otherwise,
and $t = b/\sigma$ be a primal solution. If $n' < m$, then let $u =
1/\sigma$, $v = 0$, $y_i = a_i/\sigma$ for $i = 1, \ldots , n'$ and
$y_i  = 0$ otherwise be a dual solution; if $n' = m$, let $u =
1/\sigma$, $v = a_{m}/\sigma$, $y_i = (a_i - a_{m})/\sigma$ for $i =
1, \ldots , m$ and $y_i  = 0$ otherwise be a dual solution. In either
case, we can verify that $(\bm{x}, t)$ and $(u, v, \bm{y})$ are both
feasible and have the same objective values. Consequently, $(\bm{x},
t)$ is an optimal solution to (LP2). Since $n' \leq m$ and all the
jobs have processing times either $t = b/\sigma$ or $0$, we can see
that $(\bm{x}, t)$ is feasible to the SLC problem, and hence it is
optimal. $\hfill\Box$

\section{Fixed Number of Machines ($m\geq 2$)}
\label{sec:fm}

In this section, we discuss the case where the number of machines
$m$ is at least two but is still a fixed constant. We consider two
further cases: when the number of constraints is also fixed and when
the number of constraints is an input of an instance.

\subsection{Fixed Number of Constraints ($k \geq 2$)}
\label{sec:fmfk}

We show that when both $m$ and $k$ are at least two but are still
fixed constants, the SLC problem is polynomial-time solvable. First,
we prove the following property of the SLC problem:

\begin{lemma}\label{lem:nonzero}
The SLC problem has an optimal solution in which at most $m + k - 1$
jobs have nonzero processing times.
\end{lemma}

{\noindent\bf Proof.} We prove that given any optimal solution to
the SLC problem, we can find an optimal solution that satisfies the
desired property. To show this, suppose we have an optimal solution
to the SLC problem in which $I_l$ is the set of jobs that are
assigned to machine $l$. We construct the following linear program:
\begin{eqnarray*}{\textrm{(LP3)}}\hspace{0.5cm}
\begin{array}{lcl}
\min & t  \\
s.t. & A \bm{x} \geq \bm{b}\\
& \sum\limits_{i \in I_l} x_{i} \leq t & \qquad\forall l = 1, \ldots , m\\
& \bm{x}, t \geq 0.
\end{array}
\end{eqnarray*}
It can be seen that any optimal solution to (LP3) is optimal to the
SLC problem. Note that there are totally $m + k$ linear constraints
(except for the nonnegative constraints) in (LP3), therefore each of
its basic feasible solutions has at most $m + k$ nonzero entries.
Now consider the variable $t$ in any basic feasible solution. If $t
= 0$, then all the processing times are zero and the lemma holds.
Otherwise, there are at most $m + k - 1$ nonzero $x_i$s in this
basic feasible solution. This implies that there exists an optimal
solution which has at most $m + k - 1$ nonzero processing times and
thus the lemma holds. $\hfill\Box$\\

By Lemma \ref{lem:nonzero}, there exists an optimal solution that
contains a constant number of nonzero processing times. In view of
this, we can find the optimal solution by enumeration. Our approach
is to first enumerate all the nonzero processing time jobs and fix
their assignments. Then we solve (LP3) to find the best processing
times. We denote $J$ as the job set and state the details of this
procedure in Algorithm \ref{alg:enum}:

\begin{algorithm}[htb]
\caption{Enumeration algorithm for fixed $m$ and fixed $k$}
\label{alg:enum}
\begin{algorithmic}[1]
\FOR {each subset $J'$ of $J$ with $m + k - 1$ jobs}

\FOR {each possible assignment of the jobs in $J'$ to the $m$
machines}

\STATE Solve (LP3) while setting $x_i = 0$ for $i\not\in J'$.

\IF {(LP3) is feasible} \STATE {Let the processing times of jobs be
the optimal solution to (LP3), and record the schedule and the
makespan.} \ENDIF \ENDFOR \ENDFOR \STATE {\bf return} the schedule
with the smallest makespan among all these iterations and its
corresponding processing times.
\end{algorithmic}
\end{algorithm}

\begin{theorem}
Algorithm \ref{alg:enum} returns an optimal solution to the SLC
problem and its computational complexity is $O(n^{m + k - 1}L)$.
\label{th:fmfk}
\end{theorem}

{\noindent\bf Proof.} The optimality follows from Lemma
\ref{lem:nonzero}, and the fact that there must be an assignment in
the enumeration which is identical to the assignment in the true
optimal solution. Then when one solves (LP3) with that assignment,
an optimal solution will be obtained.
%

Now we study the total running time of Algorithm \ref{alg:enum}.
There are at most $O\left(\dbinom{n}{m + k - 1}(m + k - 1)^m\right)
= O(n^{m + k - 1})$ cases in the enumeration algorithm. In each case, we need
to solve one linear program (LP3), which has $m + k$ variables
($m+k-1$ for $\bm{x}$ and $1$ for $t$) and the same number of constraints. The running time for solving
the linear program is $O((m+k)^{3}L)$. Therefore, in total,
Algorithm \ref{alg:enum} requires $O\left(n^{m + k - 1}(m + k
)^{3}L\right) = O(n^{m + k - 1}L)$
operations. $\hfill\Box$\\

We close this subsection by considering the simple cases where $m =
1$ and $k = 2$. In Section \ref{sec:1m}, we show that these cases
can be solved in $O(n^3L)$ time via solving the linear program
(LP1). Notice that using the enumeration algorithm above, the
worst-case running time in this case can be improved to $O(n^{2}L)$.

\subsection{Arbitrary Number of Constraints ($k \geq 2$)}
\label{sec:fmak}

Now we consider the case in which the number of constraints $k$ is
also an input in the problem. In this case, it is easy to see that
the classical parallel machine scheduling problem is a special case
of the SLC problem, as we can set $A$ in (\ref{eq:LP}) to be an
identity matrix and $\bm{b}$ to be the predetermined processing times of
the jobs. Therefore, the hardness result for the parallel machine
scheduling problem also stands for the SLC problem, i.e., the SLC
problem is NP-Hard when the number of machines is fixed and is
strongly NP-Hard when the number of machines is an input
\citep{GJ79}. In the following, we focus on designing approximation
algorithms for this case.

We first design a PTAS for the case where the number of machines is
fixed and the number of constraints is an input. The result is based
on guessing the optimal values of the large jobs and the PTAS for
the parallel machine scheduling problem with a fixed number of
machines by \cite{Graham69}.

Before describing our algorithm, we define $P$ to be the optimal
value of the following linear program:
\begin{eqnarray*}\label{eq:fm_lp}
\begin{array}{ll}
\min & \sum\limits_{i = 1}^n x_i  \\
s.t. & A\bm{x} \geq \bm{b} \\
     & \bm{x} \geq 0.
\end{array}
\end{eqnarray*}
Apparently, $P$ is an upper bound and $P/m$ is a lower bound of the
optimal makespan to the SLC problem. In addition, $P$ is polynomial
in the input sizes, $n$ and $k$. We use $\lceil x \rceil$ to denote
the smallest integer that is greater than or equal to $x$. The PTAS
for this case is described in Algorithm \ref{alg:ptas}.

\begin{algorithm}[h]
\caption{PTAS for fixed $m$ and arbitrary $k$}\label{alg:ptas}
\begin{algorithmic}[1]
\STATE Given $\epsilon \in (0, 1)$ and $P$ defined as before.

\STATE Let $h = \lceil (m - 1)/\epsilon\rceil$, and divide $[0, P]$ into
$T_0 = 0$, $T_1 = \epsilon P/m,\ldots,T_{i + 1} = (1 +
\epsilon)^i\epsilon P/m,\ldots,T_{l - 1} = (1 + \epsilon)^{l -
2}\epsilon P/m$, $T_{l} = P$, where $l$ is defined such that $(1 +
\epsilon)^{l - 2}\epsilon P/m < P \leq (1 + \epsilon)^{l -
1}\epsilon P/m$. \FOR {each subset $J^h$ of $J$ with $h$
jobs}\label{alg:ptas_2}

\FOR {each combination of $P_{i}\in \{T_1, \ldots , T_{l}\}$, $i\in
J^h$}\label{al:fm_2} \STATE Set $Q_{i} = T_{j-1}$ if $P_{i} =
T_{j}$, $\forall i\in J^h$ \STATE Let $J^r = J \setminus J^h$. Solve
the following linear program:
\begin{eqnarray*}{\textrm{(LP4)}}\hspace{0.5cm}
\begin{array}{lll}
\min & \sum\limits_{i = 1}^{n} x_i\\
s.t. & A\bm{x} \geq \bm{b}\\
& x_j \leq x_i & \forall j \in J^r, i \in J^h\\
& Q_{i} \leq x_i \leq P_{i} & \forall i \in J^h\\
& \bm{x} \geq 0.
\end{array}
\end{eqnarray*}
\IF {(LP4) is feasible}

\STATE Let the processing times of jobs be the optimal solution to
(LP4), and $J^0$ be the jobs in $J^h$ that have processing times in
$[T_0, T_1]$. \FOR {each possible assignment of the jobs in
$J^h\setminus J^0$ to the $m$ machines} \label{alg:ptas_opt} \STATE
Apply list scheduling to the remaining jobs in $J^0 \cup J^r$, and
record the schedule and the processing times. \ENDFOR \ENDIF \ENDFOR
\ENDFOR \STATE {\bf return} the schedule with the smallest makespan
among all these iterations and its corresponding processing times.
\end{algorithmic}
\end{algorithm}

\begin{theorem}\label{th:ptas}
Algorithm \ref{alg:ptas} is a PTAS for the SLC problem when the
number of constraints $k$ is an input of the instance, and the
number of machines $m$ is a fixed constant.
\end{theorem}

{\noindent\bf Proof.} First we calculate the computational
complexity of Algorithm \ref{alg:ptas}. Fixing $\epsilon$, Step
\ref{alg:ptas_2} requires $\dbinom{n}{h}$ enumerations. Note that
$(1 + \epsilon)^{l - 2}\epsilon P/m < P$, thus $l < \log
{\frac{m}{\epsilon}}\left/\log{(1 + \epsilon)} + 2 \right.\leq
\frac{2}{\epsilon}\log {\frac{m}{\epsilon}} + 2$, where the last
inequality follows from the fact that $\log(1 + \epsilon) \geq
{\epsilon}/{2}$ when $0 < \epsilon < 1$. Therefore, the number of
iterations in Step \ref{al:fm_2} is $l^h \leq
(\frac{2}{\epsilon}\log {\frac{m}{\epsilon}} + 2)^h$, which is
polynomially bounded by the input size. In each iteration, solving
the linear program (LP4) requires $O((n+k)^3L)$ operations. The
number of iterations in Step \ref{alg:ptas_opt} is $O(m^h)$ and the
list scheduling requires $O(n\log n)$ time. By the fact that $m$ and
$\epsilon$ are fixed constants, the total running time is polynomial
time.

Now we prove that the returned schedule has a makespan no larger
than $1 + \epsilon$ of the optimal makespan. Let $\bm{x}^*$ and
$C^*_{\max}$ be the processing times and the makespan of the true
optimal solution, respectively. We consider the iteration in
Algorithm \ref{alg:ptas} in which the jobs of $J^h$ are exactly the
$h$ largest jobs in the optimal schedule, the value $x^*_i$ falls in
$[Q_{i}, P_{i}]$ for each $i \in J^h$, and the assignment of jobs in
$J^h\setminus J^0$ is the same as those of the optimal solution,
where $J^0$ are the jobs in $J^h$ which have processing times in
$[T_0, T_1] = [0, \epsilon P/m]$.

In this iteration, the linear program must be feasible as $\bm{x}^*$
is a feasible solution to (\textrm{LP4}). Denote the processing
times and the makespan returned in this iteration by $\bm{x}$ and
$C_{\max}$, respectively. We study the last completed job $j$ of the
schedule. First, suppose that $j$ is in $J^h\setminus J^0$. Consider
the schedule in which we keep only the jobs in $J^h\setminus J^0$,
and the jobs are assigned to the same machines as the optimal
schedule. We denote $C_{\bm{x}^*}$ and $C_{\bm{x}}$ as the makespans
of the above schedule with processing times $\bm{x}^*$ and $\bm{x}$
respectively. Notice that $x_{i} \leq (1 + \epsilon) x^*_i$ for all
$i \in J^h\setminus J^0$  by the third set of constraints of (LP4),
therefore $C_{\bm{x}} \leq (1 + \epsilon) C_{\bm{x}^*}$. By Step 9
of the algorithm, the last completed job $j \in J^h\setminus J^0$
implies that the machine that job $j$ is assigned to only contains
jobs in $J^h\setminus J^0$. Therefore, $C_{\max} = C_{\bm{x}}$ in
this case and it follows that $C_{\max} = C_{\bm{x}} \leq (1 +
\epsilon)C_{\bm{x}^*} \leq (1 + \epsilon) C^*_{\max}$.

Next we consider the case in which the last completed job $j$ is in
$J^0\cup J^r$. There are two further cases. If $j \in J^0$, then we
have $x_j \leq \epsilon P/m \leq \epsilon C^*_{\max} \leq
\frac{m\epsilon}{m - 1} C^*_{\max}$, since $P/m$ is a lower bound of
the optimal makespan to the problem. If $j \in J^r$, then since $j$
is not one of the largest $h$ jobs, we must have $x_j \leq
\frac{1}{h}\sum_{i = 1}^{n} x_i$. And since $\bm{x}^*$ is feasible
to (LP4) and $\bm{x}$ is the optimal solution to (LP4), it follows
that $ x_{j} \leq \frac{1}{h}\sum_{i = 1}^{n} x_i \leq
\frac{1}{h}\sum_{i = 1}^{n} x^*_i \leq \frac{m}{h} C^*_{\max} \leq
\frac{m\epsilon}{m - 1} C^*_{\max}.$ Therefore, $x_j \leq
\frac{m\epsilon}{m - 1} C^*_{\max}$ for all job $j$ in $J^0\cup
J^r$.

Then since the jobs in $J^0 \cup J^r$ are scheduled by list
scheduling, we have
\begin{eqnarray*}
C_{\max } \leq \frac{1}{m} \sum_{i=1}^{n}x_{i}+\left(
1-\frac{1}{m}\right) x_{j} \leq \frac{1}{m} \sum_{i =
1}^{n}x^*_{i}+\left(1-\frac{1}{m}\right) x_{j} \leq \left( 1 +
\epsilon\right) C_{\max }^{\ast }
 \end{eqnarray*}
where the first inequality is because $j$ is the last job in the
schedule and we used the list scheduling rule, the second inequality
is because $x_i$s are optimal to (LP4) in that iteration, and the
last inequality is because $x_j \le \frac{m\epsilon}{m-1}
C^*_{\max}$ as discussed above.

Finally, note that the makespan returned by Algorithm \ref{alg:ptas}
cannot be worse than this schedule, thus Theorem \ref{th:ptas}
holds. $\hfill\Box$

\section{Arbitrary Number of Machines ($m\geq 2$)}
\label{sec:am}

In this section, we discuss the case where the number of machines is
an input. We first consider the case where there are two
constraints, and then look at the case with more than two
constraints.

\subsection{Two Constraints ($k = 2$)}
\label{sec:am2k}

In this section, we demonstrate that when there are only two
constraints, the SLC problem can be solved in polynomial time even
if the number of machines is an input of the instance. We start from
the following linear program, which is similar to (LP2):
\begin{eqnarray}\label{eq:2k}{\textrm{(LP5)}}\hspace{0.5cm}
\begin{array}{lcl}
\min & t  \label{eq:2k_o}\nonumber\\
s.t. & \sum\limits_{i = 1}^{n} a_{1i} x_i \geq b_1 \\
& \sum\limits_{i = 1}^{n} a_{2i} x_i \geq b_2 \\
& \sum\limits_{i = 1}^{n} x_{i} \leq mt\\
& x_{i} \leq t & \qquad\forall i = 1, \ldots , n \label{eq:2k_mach}\\
& \bm{x}, t \geq 0.
\end{array}
\end{eqnarray}

Next, we show that all basic feasible solutions of (LP5) are
feasible to the SLC problem. Then since any optimal solution to the
SLC problem is feasible to (LP5) by choosing $t$ as its makespan, we
know that the optimal basic feasible solution to (LP5) must also be
an optimal solution to the SLC problem. We start from the following
lemma:


\begin{lemma}\label{lem:2k_frac}
In any basic feasible solution of (LP5), there are at most two
variables in $\bm{x}$ satisfying $0 < x_i < t$. And it must be one
of the following cases: (a) exactly $m$ variables in $\bm{x}$
satisfying $x_i = t$ with all other $x_i = 0$; (b) exactly $m - 1$
variables in $\bm{x}$ satisfying $x_i = t$, and at most two
variables in $\bm{x}$ satisfying $0 < x_i < t$ with sum at most $t$;
or (c) no more than $m - 2$ variables in $\bm{x}$ satisfying $x_i =
t$, and at most two variables in $\bm{x}$ satisfying $0 < x_i < t$.
\end{lemma}

{\noindent\bf Proof.} If $t = 0$, then the lemma trivially holds.
Otherwise, we count the number of nonzero variables in the basic
feasible solution. We add a slack variable $z_i$ to
the constraint $x_{i} \leq t$ so that it is represented as $x_i + z_i = t$,
$\forall i = 1, \ldots , n$. For any fixed $i$, if
$x_i = 0$ or $x_i = t$, then the number of nonzeros (among $x_i$ and
$z_i$) in the equality $x_i + z_i = t$ is exactly one, otherwise it
is two. However, since there are $n + 3$ constraints in total, there
are at most $n + 3$ nonzero entries in a basic feasible solution of
which at most $n + 2$ are in $x_i$ and $z_i$. Therefore, for any
basic feasible solution, there can only be at most two indices $i
\in \{1, \ldots , n\}$ such that $0 < x_i < t$. The remainder of the
lemma follows immediately. $\hfill\Box$\\

Lemma \ref{lem:2k_frac} can be used directly to obtain an algorithm
for the SLC problem. We describe it as the LP-based algorithm
(Algorithm \ref{alg:lp2}). Notice that in Step \ref{alg:lp2_r} of
Algorithm \ref{alg:lp2}, Lemma \ref{lem:2k_frac} guarantees that the
sum of the processing times of the remaining jobs (at most two) is
at most $t$ if there is only one idle machine. Therefore, the
returned schedule is feasible and the makespan is $t$, the optimal
value of (LP5). Thus, we find an optimal solution to the SLC
problem.

\begin{algorithm}[htb]
\caption{LP-based algorithm for arbitrary $m$ and $k =
2$}\label{alg:lp2}
\begin{algorithmic}[1]
\STATE Find an optimal basic feasible solution $(\bm{x}, t)$ to the
linear program (LP5), and let $\bm{x}$ be the processing times of the jobs.
\STATE Schedule the jobs with processing times $x_i = t$
solely. \STATE For the remaining (at most two) jobs with $0 < x_i <
t$, if there is only one idle machine, assign these jobs on that
machine; otherwise, assign the jobs each on a solo machine.
\label{alg:lp2_r} \STATE{\bf Return} $\bm{x}$, $C_{\max} = t$, and
the schedule.
\end{algorithmic}
\end{algorithm}

The main computation in Algorithm \ref{alg:lp2} is to find an
optimal basic feasible solution. In \cite{KV12} (Theorem 4.16), a
technique is introduced to transform a feasible solution in a linear
program to a basic feasible solution by eliminating the inequality
constraints one by one, and in each round, it solves a linear
program, which requires $O(n^3L)$ operations. Thus the total running
time of Algorithm \ref{alg:lp2} is $O(n^{4}L)$.

\begin{theorem}
The SLC problem with arbitrary machines and $k = 2$ constraints can
be solved in $O(n^{4}L)$ by the LP-based algorithm.
\end{theorem}

Note that the LP-based algorithm can also be applied if the number
of machines is fixed. When the number of machines is large, the
performance of the LP-based algorithm is better than the enumeration
algorithm we presented in Theorem \ref{th:fmfk}, which requires
$O(n^{m+1}L)$ operations.

\subsection{Fixed or Arbitrary Number of Constraints $(k \geq 3)$}
\label{sec:amak}

As mentioned in the previous section, the SLC problem is strongly
NP-Hard if the number of constraints is an input of the instance. In
this section, we design two approximation algorithms for this case.
The first one is derived from the property of parallel machine
scheduling problems, and the other one is based on the technique of
linear programming. Notice that the approximation algorithms can
also be applied to the case where the number of constraints is fixed
and greater than two, however, the complexity of that case remains
unknown.

First, we design a simple approximation algorithm by adapting the
well-known list scheduling rule \citep{Graham66}. We first decide
the processing times by solving a specific linear program, and then
schedule the jobs via the list scheduling rule. The details are
given in Algorithm \ref{alg:mls}.

\begin{algorithm}[htb]
\caption{Modified list scheduling algorithm for arbitrary $k$ and
$m$}\label{alg:mls}
\begin{algorithmic}[1]
\STATE Solve the linear program below:
\begin{eqnarray*}{\textrm{(LP6)}}\hspace{0.5cm}
\begin{array}{lll}
\min & \frac{1}{m} \sum\limits_{i=1}^{n}x_{i}+\left(
    1-\frac{1}{m}\right) z \\
s.t. & A\bm{x} \geq \bm{b} \\
     & x_i \leq z & \forall i = 1, \ldots , n\\
     & \bm{x} \geq 0, \quad z \geq 0.
\end{array}
\end{eqnarray*}
Denote the optimal solution as $\bm{x}$ and $z$. \STATE Let $\bm{x}$
be the processing times of the jobs. Schedule the jobs by the list
scheduling rule.

\STATE {\bf Return} $\bm{x}$ and $C_{\max}$.
\end{algorithmic}
\end{algorithm}

We prove that the modified list scheduling algorithm has the same
approximation ratio for the SLC problem as for the classical
parallel machine scheduling problem.
\begin{theorem}
The modified list scheduling algorithm is a
($2-\frac{1}{m}$)-approximation algorithm for the SLC problem.
\end{theorem}
{\noindent\bf Proof.} The running time for solving the linear
program is $O((n+k)^{3}L)$, and for list scheduling is $O(n\log m)$.
Therefore the total running time is $O((n+k)^{3}L+n\log m)$, which
is polynomial in the input size.

Let $\bm{x}^*$, $x^*_{\max}$ and $C^*_{\max}$ be the processing
times, the maximum of the processing times and the makespan of an
optimal schedule, respectively. Let $\bm{x}$, $x_{\max}$, and
$C_{\max}$ be those returned by Algorithm \ref{alg:mls}. Consider
the last completed job $j$. By the list scheduling rule, we have
\begin{eqnarray*}
C_{\max } &\leq& \frac{1}{m} \sum_{i=1}^{n}x_{i}+\left(
1-\frac{1}{m}\right) x_{j} \leq \frac{1}{m}
\sum_{i=1}^{n}x_{i}+\left(1-\frac{1}{m}\right) x_{\max}\\
& \leq& \frac{1}{m} \sum_{i=1}^{n}x^*_{i}+\left(
1-\frac{1}{m}\right) x^*_{\max} \leq \left( 2-\frac{1}{m}\right)
C_{\max }^{\ast }.
\end{eqnarray*}

The second last inequality holds since the linear program (LP6)
returns the minimum value of such an objective function. And this bound is tight
from the tight example of the list scheduling rule for the classical
problem. $\hfill\Box$\\

The second approximation algorithm is based on the idea presented in
Section \ref{sec:am2k}. However, it is not clear how to directly
extend (LP5) to obtain an optimal solution in polynomial time. To
see this, consider a simple example with $n = 4$, $m = 3$, and $k =
3$, and the constraints are $x_1 + x_2 = x_1 + x_3 = x_1 + x_4 = 5$.
If we try to generalize (LP5), we will get the following linear
program:
\begin{eqnarray*}
\min & t \nonumber\\
s.t. & x_1 + x_2  =  x_1 + x_3 = x_1 + x_4 = 5\\
     & x_1 + x_2 + x_3 + x_4 \leq 3t\\
     & 0\le x_1, x_2, x_3, x_4 \leq t.
\end{eqnarray*}

It can be verified that the unique optimal solution to this linear
program is $(x_1, x_2 , x_3, x_4, t) = (3, 2, 2, 2, 3)$. However,
the jobs with processing times $x_2$, $x_3$, $x_4$ can not be
assigned to the two remaining machines with makespan not exceeding
three. Therefore, (LP5) may not give a feasible solution to the SLC
problem in this case. In the following, we modify it to derive an
approximation algorithm.

We consider the following linear program:
\begin{eqnarray*}{\textrm{(LP7)}}\hspace{0.5cm}
\begin{array}{lcl}
\min & t  \\
s.t. & A\bm{x} \geq \bm{b} \\
& \sum\limits_{i = 1}^{n} x_{i} \leq \left(m - K\right) t \\
& x_{i} \leq t & \qquad\forall i = 1, \ldots , n\\
& \bm{x} \geq 0,
\end{array}
\end{eqnarray*}
where $K$ is defined as:
\begin{eqnarray*}
K = \left\{\begin{array}{ll}
m - \frac{k}{k + 1 - m},& \text{if~} \tilde{k} > m, \label{eq:K1}\\
\max\left( \lceil \tilde{k}\rceil - \frac{k}{k + 1 - \lceil
\tilde{k}\rceil}, \lfloor \tilde{k}\rfloor - \frac{k}{k + 1 -
\lfloor \tilde{k}\rfloor} \right), & \text{if~} \tilde{k} \leq m,
\label{eq:K2}
\end{array}
\right.
\end{eqnarray*}
where $\tilde{k} = k + 1 - \sqrt{k}$ and $\lfloor x\rfloor$ is the
largest integer that is less than or equal to $x$.

Notice that $K = 0$ for $k = 1, 2$ and $0 < K \leq \min\{m, k\}$ for
$k \geq 3$ (see Table \ref{tab:val} for various values of $K$), and
$K$ is a rational number since $k$ and $m$ are integers. When $k =
2$, (LP7) reduces to (LP5), as considered in Section \ref{sec:am2k}.
If $k \geq 3$, however, the optimal solution to the SLC problem may
not be feasible to this linear program. Therefore, the optimal
solution to (LP7) may not be an optimal solution to the SLC problem.
Nevertheless, we prove that the optimal solution to (LP7) is still
feasible and has an objective value no larger than a factor of the
optimal makespan.

\begin{table}[htb]
\begin{minipage}[b]{1\textwidth}
\centering
\begin{tabular}{p{2em}<{\centering}|p{7em}<{\centering}|p{4em}<{\centering}|p{4em}<{\centering}|p{4em}<{\centering}|p{4em}<{\centering}}
                 \toprule
                   \multirow{2}{*}{$k$} & \multirow{2}{*}{{$\tilde{k} = k + 1 - \sqrt{k}$}} & \multicolumn{2}{c|} {$K$}& \multicolumn{2}{c} {ratio $m/(m - K)$}\\
                  \cline{3-6}
                   & & $m = 100$ & $m = 10$ & $m = 100$ & $m = 10$\\
                  \hline
                   1 & 1 & 0 & 0 & 1 & 1\\
                   2& 1.59 & 0 & 0 & 1 & 1\\
                   3 & 2.27 & 0.50 & 0.50 & 1.0050  & 1.0526\\
                   4 & 3 & 1 & 1 & 1.0101 &  1.1111\\
                   5 & 3.76 & 1.50 & 1.50 &  1.0152 & 1.1765\\
                   10 & 7.84 & 4.67 & 4.67 &  1.0490 & 1.8750\\
                   20& 16.53 & 12 & 8.18 & 1.1364 & 5.5000\\
                   50 & 43.93 & 36.86 & 8.78 & 1.5837 & 8.2000\\
                   100 & 91 & 81 & 8.90 &  5.2632 &   9.1000\\
                  \bottomrule
\end{tabular}
\caption{Values of $K$ and the Approximate Ratios for Different $k$
and $m$} \label{tab:val}
\end{minipage}
\end{table}

Similar to the case where $k = 2$, we have the following property:
\begin{lemma}\label{lem:kc_frac}
There are at most $k$ variables in $\bm{x}$ satisfying $0 < x_i < t$
for any basic feasible solution of the linear program (LP7).
\end{lemma}

We omit the proof which is analogous to that of Lemma
\ref{lem:2k_frac}.

\begin{lemma}\label{lem:kc}
For each basic feasible solution of the linear program (LP7) with $t
> 0$, if it has exactly $l$ variable(s) in $\bm{x}$ satisfying $x_i
= t$, where $l \in\{ m - \lceil K \rceil, m - \lceil K \rceil - 1,
\ldots , \max\{m - k, 0\}\}$, then it has additionally at most $k$
variables in $\bm{x}$ with each processing time $0 < x_i < t$ and
their total processing time is at most $\frac{k}{k + 1 - m + l}t$.
\end{lemma}

{\noindent\bf Proof.} Notice that there are at most $m - \lceil K
\rceil$ variables in $\bm{x}$ satisfying $x_i = t$ by the second set
of constraints of (LP7), and at most $k$ variables in $\bm{x}$
satisfying $0 < x_i < t$ by Lemma \ref{lem:kc_frac}. If there are
exactly $l$ variable(s) in $\bm{x}$ satisfying $x_i = t$, then the
sum of the remaining (at most) $k$ variables in $\bm{x}$ satisfying
$0 < x_i < t$ is no larger than $\left(m - l - K\right)t$. It
remains to show that $m - l - K$ is no greater than $\frac{k}{k + 1
- m + l}$.

For this, we define a function $f(x) = x - \frac{k}{k + 1 - x}$ for
$x\in \{1, \ldots , k\}$. It is easy to see that $f(x)$ is increasing
when $x \leq \tilde{k}$ and decreasing when $x \geq \tilde{k}$.
Therefore, when $\tilde{k} > m$, we have $m - l - K = m - l - f(m)
\leq m - l - f(m - l) = \frac{k}{k + 1 - m + l}$; when $\tilde{k}
\leq m$, we have $m - l - K = m - l - \max\limits_{x\in \{1, \ldots ,
k\}}f(x)\leq m - l - f(m - l) = \frac{k}{k + 1 - m + l}$. Therefore,
the lemma holds. $\hfill\Box$

\begin{lemma}\label{lem:kc_tran}
Let $(\bm{x}, t)$ be a basic feasible solution of (LP7), then we can
construct a feasible schedule of the SLC problem with a makespan of
at most $t$.
\end{lemma}

{\noindent\bf Proof.} If $k \leq m$ and this basic feasible solution
has at most $m - k$ variables in $\bm{x}$ satisfying $x_i = t$
(those jobs must be assigned solely), then by Lemma
\ref{lem:kc_frac}, there are at most $k$ variables in $\bm{x}$
satisfying $0 < x_i < t$ and those corresponding jobs can also be
assigned solely and we are done.

Next, we consider the case where $m < k$ or $m\ge k$ but there are
more than $m - k$ variables in $\bm{x}$ such that $x_i = t$. Let
$\l$ denote the number of variables in $\bm{x}$ such that $x_i = t$.
By the case assumption and the constraint in (LP7), $l \in \{m -
\lceil K \rceil, m - \lceil K \rceil - 1, \ldots , \max\{m - k, 0\}\}$.
Without loss of generality, we can assume there are exactly $k$ jobs
having processing times $x_i < t$, with some jobs having possibly zero
processing times. We now show how to construct a feasible schedule
with a makespan at most $t$. First, we assign the $l$ jobs with
processing times $x_i = t$ to $l$ machines solely. Then we find the
smallest $k + 1 - m + l$ jobs with processing times smaller than
$t$. We claim that these jobs have a total processing time of at
most $t$, and hence can be fit into a single machine. If not, it
follows that any $k + 1 - m + l$ of the $k$ jobs with processing
times smaller than $t$ have a total processing time greater than
$t$. Then we have the following $k$ inequalities (mod $k$ for each
index):
\begin{eqnarray}\label{eq:kc_fea}
\begin{array}{c}
x_1 + x_2 + \cdots + x_{k + 1 - m + l}  > t, \\
x_2 + x_3 + \cdots + x_{k + 2 - m + l}  >  t,\\
\vdots\\
x_{k - 1} + x_k + \cdots + x_{k - 1 - m + l}  >  t,\\
x_{k} + x_1 + \cdots + x_{k - m + l}  >  t.
\end{array}
\end{eqnarray}
On the one hand, summing up inequalities in (\ref{eq:kc_fea}) we
obtain $(k + 1 - m + l) (x_1 + x_2 + \cdots + x_k) > kt$, or $x_1 +
x_2 + \cdots + x_k  > \frac{k}{k + 1 - m + l}t$. On the other hand,
by Lemma \ref{lem:kc}, the total processing time of the jobs with
processing times $x_i < t$ is at most $\frac{k}{k + 1 - m + l}t$,
which leads to a contradiction. Finally, there are at most $k - (k +
1 - m + l) = m - l - 1$ jobs each with processing time smaller than
$t$, and $m - l - 1$ remaining machines. Assigning these jobs solely
provides a feasible
solution with a makespan of at most $t$. $\hfill\Box$\\

Similar to LP-based algorithm for $k = 2$, we can find an
optimal basic feasible solution of (LP7) in polynomial time and
obtain a feasible approximated schedule for the SLC problem. We also
call it LP-based algorithm and summarize it in Algorithm
\ref{alg:lpk}. Next we study its approximation ratio.

\begin{algorithm}[htb]
\caption{LP-based algorithm for general SLC}\label{alg:lpk}
\begin{algorithmic}[1]
\STATE Find an optimal basic feasible solution $(\bm{x}, t)$ of the
linear program (LP7), and let $\bm{x}$ be the processing times of
the jobs. \STATE Let $l$ be the number of jobs with processing times
$x_i = t$. Schedule these jobs solely. If there is any remaining
job, continue the following process. \STATE Find the smallest $k + 1
- m + l$ jobs with processing time $0 < x_i < t$, and schedule them
in a single machine.\label{alg:lpk_3} \STATE Schedule the remaining
jobs in the remaining $m - l - 1$ machines solely. \STATE {\bf
Return} $\bm{x}$, $C_{\max} = t$, and the schedule.
\end{algorithmic}
\end{algorithm}

\begin{theorem}\label{lem:kc_app}
The schedule returned by the LP-based algorithm has a makespan of
$C_{\max} = t \leq \frac{m}{m - K}C^*_{\max}$, where $C^*_{\max}$ is
the optimal makespan of the problem.
\end{theorem}

{\noindent\bf Proof.} Let $\bm{x}^*$ be the optimal schedule with
makespan $C_{\max}^*$. It suffices to show that $(\bm{x}^*,
\frac{m}{m - K}C^*_{\max})$ is a feasible solution to (LP7). If that
holds, then since Algorithm \ref{alg:lpk} provides a schedule with
makespan $t$, the optimal value of (LP7), the results hold. To show
that, note that $\bm{x}^*$ satisfies the first set of linear
constraints of (LP7). $C^*_{\max}$ is the optimal makespan implies
that $\sum\limits^n_{i = 1} x^*_i \leq mC^*_{\max} = (m - K)
\frac{m}{m -  K} C^*_{\max}$, and $x^*_i \leq C^*_{\max} \leq
\frac{m}{m - K}C^*_{\max}$ for each $i$. Therefore, $(\bm{x}^*,
\frac{m}{m - K}C^*_{\max})$ is a feasible solution. Thus the theorem
holds. $\hfill\Box$\\

We show some examples of the ratios $m/(m - K)$ in Table
\ref{tab:val}. We can see that the LP-based algorithm performs well
when $k$ is relatively small. The performance deteriorates when $k$
becomes large. Furthermore, if $k$ is a fixed constant, the
approximate ratio is close to one when $m$ is sufficiently large.

Finally, we note that in practice one can apply both approximation
algorithms \ref{alg:mls} and \ref{alg:lpk}, and choose the one that
gives a better solution. The approximation ratio of the combined
algorithm will be $\min\{\frac{m}{m-K}, 2-\frac{1}{m}\}$.

\section{Conclusion}
\label{sec_end}

In this paper, we present a scheduling problem in which the
processing times of jobs satisfy a set of linear constraints. We
discuss the computational complexity and propose several algorithms
for the problem. There are two open questions left: whether the case
with a fixed number of constraints (more than two) and an arbitrary
number of machines is polynomial-time solvable or not (Section
\ref{sec:amak}), and how to design an FPTAS or prove that it does
not exist for the case with an arbitrary number of constraints and a
fixed number of machines (Section \ref{sec:fmak}).

The current research may inspire further explorations about the
scheduling problems with other machine environments, such as
uniformly related parallel scheduling, shop scheduling problems, or
under various restrictions, e.g., precedences and release dates, or
minimizing other objective criterion. One can also consider other
combinatorial optimization problems under linear constraints.

\section*{Acknowledgments}
Zhenbo Wang's research has been supported by NSFC No. 11371216.

\section*{References}
\bibliographystyle{model5-names}\biboptions{authoryear}
\bibliography{ref_slc_r1}

\end{document}